\documentclass[conference]{IEEEtran}
\IEEEoverridecommandlockouts

\usepackage{cite}
\usepackage{amsmath,amssymb,amsfonts}
\usepackage{algorithmicx}
\usepackage{algpseudocode}
\usepackage{graphicx}
\usepackage{tabularx} 
\usepackage{textcomp}
\usepackage{xcolor}
\usepackage{float}
\usepackage{amsmath}
\usepackage{mathrsfs}
\usepackage{enumitem}
\usepackage{booktabs}
\def\BibTeX{{\rm B\kern-.05em{\sc i\kern-.025em b}\kern-.08em
    T\kern-.1667em\lower.7ex\hbox{E}\kern-.125emX}}
\begin{document}

\title{Efficient Speech Watermarking for Speech Synthesis via Progressive Knowledge Distillation}
\author{
\IEEEauthorblockN{Yang Cui}
\IEEEauthorblockA{\textit{Affiliation}\\
Beijing, China\\
Author1@Affiliation.com}
\and
\IEEEauthorblockN{Author2}
\IEEEauthorblockA{\textit{Affiliation}\\
Beijing, China\\
Author1@Affiliation.com}
\and
\IEEEauthorblockN{Author3}
\IEEEauthorblockA{\textit{Affiliation}\\
Beijing, China\\
Author1@Affiliation.com}
\and
\IEEEauthorblockN{Author4}
\IEEEauthorblockA{\textit{Affiliation}\\
Beijing, China\\
Author1@Affiliation.com}
}

\author{
\IEEEauthorblockN{Yang Cui \quad Peter Pan \quad Lei He \quad Sheng Zhao}
\IEEEauthorblockA{Microsoft, Beijing, China\\
\{yancu, peterpan, helei, szhao\}@microsoft.com}
}

\maketitle

\begin{abstract}
With the rapid advancement of speech generative models, unauthorized voice cloning poses significant privacy and security risks. Speech watermarking offers a viable solution for tracing sources and preventing misuse. Current watermarking technologies fall mainly into two categories: DSP-based methods and deep learning-based methods. DSP-based methods are efficient but vulnerable to attacks, whereas deep learning-based methods offer robust protection at the expense of significantly higher computational cost. To improve the computational efficiency and enhance the robustness, we propose PKDMark, a lightweight deep learning-based speech watermarking method that leverages progressive knowledge distillation (PKD). Our approach proceeds in two stages: (1) training a high-performance teacher model using an invertible neural network-based architecture, and (2) transferring the teacher's capabilities to a compact student model through progressive knowledge distillation. This process reduces computational costs by 93.6\% while maintaining high level of robust performance and imperceptibility. Experimental results demonstrate that our distilled model achieves an average detection F1 score of 99.6\% with a PESQ of 4.30 in advanced distortions, enabling efficient speech watermarking for real-time speech synthesis applications.
\end{abstract}


\section{Introduction}
With the rapid advancement of deep learning-based generative models, text-to-speech (TTS) and voice cloning technologies have reached unprecedented levels of realism. While these innovations enable high-fidelity speech synthesis, they also introduce significant challenges in copyright protection, content authentication, and privacy security. The increasing difficulty in distinguishing synthetic speech from natural speech raises concerns about misinformation, deepfake attacks, and unauthorized content reproduction.

Audio watermarking has emerged as a proactive and effective solution by embedding traceable, imperceptible signals within speech, enabling robust identification and authentication \cite{hua2016twenty}.
Traditional audio watermarking methods relied on digital signal processing (DSP) to embed watermarks in time domain or transformation domains, demonstrating resilience against common distortions such as noise addition or lossy compression. However, their reliance on handcrafted features makes them susceptible to advanced manipulations, including reverberation and desynchronization attacks \cite{liu2018patchwork}.

To overcome these limitations, researchers have explored deep learning-based methods that enhance the imperceptibility, robustness, and capacity of audio watermarks. These methods generally fall into two categories: generative watermarking and post-hoc watermarking.

Generative watermarking techniques directly embed watermarks into the speech synthesis process, ensuring that all generated speech inherently carries a watermark. While this approach eliminates the need for additional processing, the embedded watermark is model-specific and lacks adaptability across different synthesis models. For example, TraceableSpeech \cite{zhou2024traceablespeech} embeds watermarking into a TTS model, ensuring high imperceptibility while remaining resilient to audio clipping attacks. Similarly, \cite{san2025latent} introduces a method for watermarking audio generative models at the latent representation level, ensuring robust detection. To enhance flexibility, HiFi-GANw \cite{cheng2024hifi} fine-tunes a HiFi-GAN vocoder \cite{kong2020hifigan} by jointly optimizing speech quality and watermark extraction loss, enabling robust watermark retrieval across various synthesis models. Building on the collaborative watermarking framework, \cite{juvela2025audio} enhances the robustness to both traditional and neural audio codecs by employing a HiFi-GAN vocoder and an anti-spoofing detector. In another generative framework, GROOT \cite{liu2024groot} embeds watermarks during audio synthesis using diffusion models, achieving high extraction accuracy even under compound post-processing attacks. 
\begin{figure*}[t]
  \centering
  \includegraphics[width=\linewidth]{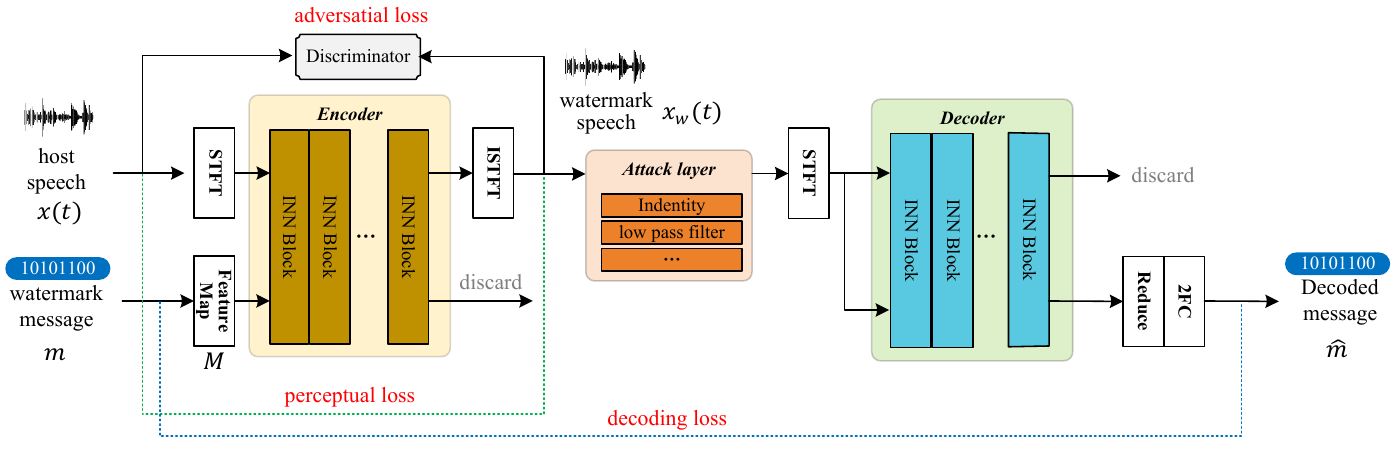}
  \caption{Overview of the Proposed Watermark Framework.}
  \label{fig:overviewframework}
\end{figure*}

On the other hand, post-hoc watermarking applies watermarking techniques after speech synthesis, offering flexible integration but requiring additional processing. As a representative example, WavMark \cite{chen2023wavmark} introduces an invertible neural network (INN)-based approach, allowing flexible watermark embedding and detection with high robustness. Extending this approach, DRAW \cite{li2024draw} enhances resilience against desynchronization and replay attacks in a dual-decoder framework. While early approaches mainly ensured basic robustness, recent work increasingly targets advanced manipulations. DeAR \cite{liu2023dear} and DeepAWR \cite{lin2025audio} employ deep learning to improve the robustness against re-recording distortions. To further mitigate manipulation attacks, a method proposed in \cite{wen2024robust} adopts an encoder-decoder framework in the discrete wavelet transform (DWT) domain. Additionally, a multi-scale transformer model \cite{tong2024enhancing} extracts acoustic features to strengthen resistance against desynchronization. MaskMark \cite{o2024maskmark} embeds a secret key vector through a multiplicative spectrogram mask to improve robustness against neural network-based transformations. Along with efforts on robustness, efficiency has also been explored. AudioSeal \cite{roman2024proactive} employs an EnCodec\cite{defossez2023high}-based generator–detector architecture for fast, localized watermark detection at the sample level. By integrating psychoacoustic masking with a compact model design, SilentCipher \cite{singh2024silentcipher} enhances message capacity and robustness while maintaining imperceptibility. Operating on discrete latent representations, DiscreteWM \cite{ji2025speech} employs vector-quantized autoencoders to embed watermarks, enhancing robustness while offering flexible capacity and efficient detection.

Deep learning–based speech watermarking offers strong robustness but often comes with high computational costs, limiting its practicality in real-time applications. In contrast, traditional DSP-based watermarking is computationally efficient but lacks resilience against sophisticated attacks. To enhance both efficiency and robustness, we propose PKDMark, a Progressive Knowledge Distillation (PKD) framework that combines the robustness of deep models with the efficiency of lightweight architectures, enabling practical deployment in real-world speech synthesis and transmission systems. The key contributions of the paper are summarized as follows: 

(1) The robustness of an INN-based speech watermarking system against advanced signal distortions is enhanced by introducing a complex domain message feature map and a multi-objective optimization strategy that jointly improves perceptual quality and decoding accuracy.

(2) A PKD-based method is introduced to transfer the watermarking capability from a teacher model with strong robustness and high perceptual quality. Through progressive distillation, the lightweight student model incrementally learns these capabilities, enabling efficient real-time watermark embedding.

(3) Furthermore, the proposed framework exhibits strong scalability and generalizability, offering a viable solution for a broad class of deep learning–based post hoc watermarking systems beyond speech synthesis.

\section{Related Work}
\subsection{WavMark}
WavMark \cite{chen2023wavmark} is a deep learning-based audio watermarking framework built on an invertible neural network (INN), enabling shared parameters for both watermark embedding and decoding. It encodes up to 32 bits per second while maintaining high imperceptibility and a low average bit error rate (BER) of 0.48\% across ten types of attacks. WavMark further supports automatic localization, allowing precise watermark detection without external reference. Its robustness and domain versatility make it well-suited for voice cloning detection, copyright protection, and broadcast authentication.

\subsection{Knowledge Distillation}
Knowledge distillation (KD) \cite{hinton2015distilling} is a well-known neural model compression method where a smaller student model is generally supervised by a larger teacher model to obtain a competitive performance. 
Progressive knowledge distillation (PKD) \cite{gou2021knowledge} is an extension of traditional knowledge distillation, designed to enhance the knowledge transfer process from teacher to student. Instead of transferring all knowledge at once, PKD gradually refines the student model over intermediate learning stages. This approach has been demonstrated to be effective in improving the performance of large language model \cite{xu2024survey}, image classification and generation \cite{shi2021follow, rezagholizadeh2022pro, Salimans2022Progressive} and information retrieval \cite{lin2023prod}. Different from prior PKD studies, the proposed PKD approach explores knowledge distillation in the context of speech watermarking for the first time. By jointly optimizing both the teacher and student models, it enables more effective knowledge transfer and improves the student model’s ability to embed robust watermarks.
\section{Method}
\subsection{Overview of framework}
The proposed method adopts an \textit{encoder–attack–decoder} pipeline to enable robust speech watermarking, where an intermediate attack layer simulates real-world distortions to enforce resilience against manipulations. An overview of the proposed method framework is shown in Figure~\ref{fig:overviewframework}. The process begins by converting the input host speech signal $x(t)$ into a complex-valued spectrogram $X \in \mathbb{C}^{T \times F}$ via Short-Time Fourier Transform (STFT), where $T$ denotes the number of time frames and $F$ the number of frequency bins. A binary message \(m\in\{0,1\}^K\) of length $K$ is mapped and expanded to a complex feature map $M\in\mathbb{C}^{T\times F}$. The encoder network fuses \(X\) and \(M\) to generate a watermarked spectrogram $X_w$, which is subsequently transformed back to the time domain using inverse STFT, yielding the watermarked speech signal $x_w(t) = \mathrm{ISTFT}(X_w)$. To simulate real‐world distortions, an attack layer applies various of distortions before message decoding, ensuring watermark resilience against different types of manipulations. The distorted speech $x_{a}(t)$ is then transformed into a complex STFT spectrogram $\hat{X}_a$ and passed through the decoder to recover a decoded feature map $\hat{M}$. Finally, the decoded message $\hat{m}$ is obtained by reducing $\hat{M}$ along the temporal dimension and feeding the reduced vector through two feedforward layers.

Building on this foundation, our main contribution is a two‐stage training strategy that enhances robustness, imperceptibility, and efficiency. In the first stage, a teacher model is trained for end-to-end audio watermarking using an invertible neural network (INN), guided by a multi-objective optimization scheme. In the second stage, a lightweight student model is trained via progressive knowledge distillation (PKD), where supervision gradually transitions from the teacher’s guidance to the student’s own predictions. This design enables efficient watermark embedding while preserving the strong robustness and high perceptual quality of the teacher watermarking system. For watermark detection, the teacher decoder is retained to ensure accurate and reliable message recovery.

\subsection{Teacher Model}
The teacher model is trained by jointly optimizing the encoder and decoder to balance imperceptibility and robustness, with an adversarial discriminator further enhancing speech fidelity. Compared to WavMark \cite{chen2023wavmark}, our approach introduces a novel message feature map in the complex STFT domain, along with a multi-objective joint optimization strategy. By exploiting both amplitude and phase information, it alleviates distortions caused by phase–amplitude mismatches during ISTFT reconstruction. The message feature map embeds information in both the real and imaginary components, ensuring alignment with the time and frequency structures in the complex STFT domain of the original speech. Additionally, the multi-objective optimization ensures an optimal balance between imperceptibility and robustness, resulting in superior performance of the teacher model compared to WavMark.

\subsubsection{Model Architecture}
The teacher model utilizes an invertible neural network (INN) \cite{chen2023wavmark} in which the encoder and decoder share parameters, ensuring strict reversibility in both forward and inverse processes. Training procedure follows an encoder-attack-decoder framework, incorporating a signal processor, message feature mapping, encoder, decoder, attack layer, and discriminator. The detailed modules and training flow are as follows.

\begin{enumerate}[label=\roman*., align=left, left=0pt, labelwidth=0pt, itemindent=0pt, labelsep=0.5em, listparindent=0pt, wide=0pt, nosep]
    \item \textbf{Signal processor:} 
    Input speech is converted to the STFT domain using a sliding window and FFT analysis, producing a complex representation. The watermarked STFT spectrogram is transformed back to the time domain using ISTFT.
    \item \textbf{Message Feature Map:} 
    An embedding layer maps each bit to a corresponding message embedding vector. A two-layer feedforward neural network transforms the embedding vectors to align with the frequency dimension $F$ of complex STFT spectrogram $X$ of the host speech. The hidden vectors are then summed and averaged, forming a compact representation $\mathbf{h}$.  Finally, the transformed representation is repeated along the time and complex axes, ensuring it matches the time-frequency dimension of $X$, resulting in a complex message feature map.
\begin{equation}
\mathbf{M} = \mathrm{Repeat}_T\left( \mathbf{h} + j \cdot \mathbf{h} \right)
\end{equation}

where $\mathbf{M} \in \mathbb{C}^{T \times F}$ is the complex-valued message feature map aligned with the shape of the host speech STFT. The vector $\mathbf{h} \in \mathbb{R}^{F}$ represents the average output of a two-layer feedforward neural network applied to the message embeddings. The operator $\mathrm{Repeat}_T(\cdot)$ denotes repeating and concatenating the vector $\mathbf{h}$ along the time axis $T$ times. Here, $j = \sqrt{-1}$ is the imaginary unit, and the same vector $\mathbf{h}$ is used as both the real and imaginary part. This design enables the message embedding to jointly leverage both the real and imaginary parts of the complex-valued representation.
    \item \textbf{Encoder:} 
    The encoder receives the complex STFT spectrogram of the original audio and the message feature map, transforming them through a few INN blocks, resulting in the complex watermarked STFT spectrogram.
    \item \textbf{Decoder:} 
    The decoder performs the inverse process of the encoder. For simplicity, it takes the watermarked STFT as dual input to extract the watermark feature map. A reduction operation averages the hidden vectors across the time and complex axes. Finally, a two-layer feedforward neural network reconstructs the extracted watermark bits from the decoded message feature map.
    \item \textbf{Attack Layer:} 
    Fourteen types of distortions (detailed in Section IV.C) are introduced to interfere with the watermarked signal, comprising nine basic distortions and five advanced attacks. These attacks simulate practical conditions, including degradation, cropping, and noise, ensuring the watermark remains robust against basic and advanced challenges.
    \item \textbf{Discriminator:} The discriminator is to distinguish the watermarked speech and the original speech, providing adversarial feedback to the entire network by computing a cross-entropy loss. 
\end{enumerate}
\subsubsection{Multi-Objective Joint Optimization}
During the training of the teacher model, multiple loss terms are introduced and jointly optimized to balance imperceptibility, robustness, and decoding accuracy:
\begin{itemize}[left=0pt, itemindent=0pt, labelsep=0.5em, listparindent=0pt, align=left, wide=0pt, nosep]
    \item \textbf{Perceptual Loss}: 
     A multi-scale STFT loss \cite{roman2024proactive} is adopted to measure the difference in the time-frequency domain between the watermarked speech and the original speech. Specifically, magnitude spectral differences are computed at various window sizes and frequency resolutions and then combined via weighted summation. The perceptual loss $\mathcal{L}_{\text{per}}$ can be formulated as:
\begin{equation}
\mathcal{L}_{\text{per}} = \sum_{i \in e} \alpha_i \|STFT_i(x(t)) - STFT_i(x_w(t))\|_2
\end{equation}

where $x(t)$ represents the original speech, $x_w(t)$ represents watermarked speech. $STFT_i$ is a normalized STFT with a window size of $2^i$ and a hop length of $2^{i/4}$. $e = 5, \ldots, 11$ is the set of scales and $\alpha$ represents the set of scalar coefficients that balance the terms.

    \item \textbf{Decoding Loss}:
    The MSE loss is used to measure the difference between the decoded message and the original message. Additionally, another MSE loss is computed between the original and decoded message feature maps to ensure fine-grained alignment in the time-frequency domain, thereby ensuring accurate recovery of the embedded information. The decoding loss $\mathcal{L_\text{dec}}$ is defined as:
    \begin{equation}
\mathcal{L}_{\text{dec}} = \| m - \hat{m} \|_2 + \alpha \| \mathbf{M} - \hat{\mathbf{M}} \|_2
\end{equation}

where $m$ denotes the embedded message and $\hat{m}$ is the decoded message. $\mathbf{M}$ represents the message feature map in the encoder, while $\hat{M}$ denotes the corresponding feature map predicted in the decoder. The scalar $\alpha$ is a balancing factor that controls the contribution of the second item.

    \item \textbf{Adversarial Loss}:
    The adversarial loss is computed via the discriminator, which attempts to differentiate between watermarked audio $x_w(t)$ and the original audio $x$. The resulting cross-entropy loss drives the adversarial training process, thereby improving the fidelity of the watermarked audio.
    \begin{equation}
    \mathcal{L}_{dis} = \log(1 - d(x(t)) + \log(d(x_w(t)))
    \end{equation}
    \begin{equation}
    \mathcal{L}_{adv} = \log(1 - d(x_w(t)))
    \end{equation}
where $\mathcal{L}_{dis}$ and $\mathcal{L}_{adv}$ denote the loss functions of the discriminator and the encoder, respectively. $d(\cdot)$ denotes the output of the discriminator, representing the probability that the input is identified as watermarked audio.

\end{itemize}
The total loss $\mathcal{L_\text{total}}$ is computed as a weighed sum of the above loss terms:
\begin{equation}
    \mathcal{L_\text{total}} = \lambda_{\text{per}} \mathcal{L}_{\text{per}} + \lambda_{\text{dec}}\mathcal{L}_{\text{dec}} + \lambda_{\text{adv}} \mathcal{L}_{\text{adv}} \text{,}
\end{equation}
where \( \lambda_{\text{per}} \), \( \lambda_{\text{dec}} \), and \( \lambda_{\text{adv}} \) are the corresponding weights of the perceptual loss \( \mathcal{L}_{\text{per}} \), the decoding loss \( \mathcal{L}_{\text{dec}} \) and the adversarial loss \( \mathcal{L}_{\text{adv}} \).

\subsection{Student Model}
To reduce model size and improve inference efficiency, direct knowledge distillation (DKD) is initially explored by jointly training the student encoder with the teacher decoder. However, experiments (detailed in Section IV.E) show that this approach falls short of the teacher model's performance, as it primarily transfers knowledge at the output layer while overlooking feature representation learning. To address this, we propose a progressive knowledge distillation (PKD) method in training, which gradually transferrs watermark knowledge from the teacher to the student model.
\subsubsection{Progressive Knowledge Distillation}
The overall PKD framework is illustrated in Figure 2. Specifically, the student encoder output $x_w^S(t)$ and the teacher encoder output $x_w^T(t)$ are linearly combined as follows:
\begin{equation}
    x_{\text{com}}(t) = \lambda(n) \cdot x_w^S(t) + (1 - \lambda (n)) \cdot x_w^T(t), 
\end{equation} 
where \( \lambda(n) \) denotes the mixing factor of student model, with n representing the training step. The combined output $X_{\text{com}}$ is decoded by the teacher decoder. During training, \( \lambda(n) \) gradually increases from 0 to 1. Initially, When \( \lambda(n) \) = 0, the training framework is equivalent to the teacher model training framework. Conversely, when \( \lambda(n) \) = 1, it corresponds to the direct knowledge distillation framework.

Adopting a progressive strategy to increase \( \lambda(n) \) enables the student model to strike a better balance between perceptual quality and decoding robustness. In the early training stage, a smaller \( \lambda(n) \) ensures that gradient updates are primarily guided by the teacher network, providing a stable supervisory signal. This guidance helps the student network build a strong initial representation and prevents it from getting trapped in local optima.
As training progresses and the student network improves, gradually increasing \( \lambda(n) \) shifts the gradient updates to rely more on the student’s own outputs. This transition encourages the student model to improve its decoding performance by optimizing its own representations while still benefiting from subtle guidance from the teacher network. In the later stage, when \( \lambda(n) \) = 1, the student model has effectively learned the teacher’s representations and can be jointly optimized with the teacher decoder to further enhance performance.
\begin{figure}[t]  
  \centering
  \includegraphics[width=\linewidth]{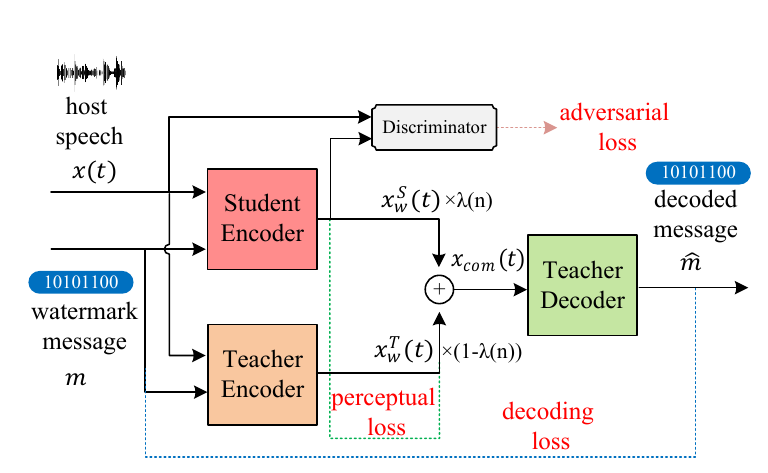}
  \caption{Overview of the PKD Framework.}
  \label{fig:overviewpkd}
\end{figure}

\begin{table*}[t]
    \centering
    \caption{Performance comparison with test speech watermarking methods. Best performance in each column is highlighted in bold.}
    \label{tab:comparison}
    \resizebox{\textwidth}{!}{%
    \begin{tabular}{lccc*{15}{c}}
        \toprule
        Models & BPS($\uparrow$) & PESQ($\uparrow$) & SNR(dB)($\uparrow$) & \multicolumn{15}{c}{BER(\%)($\downarrow$)} \\
               &                 &                 &                 & MEAN & ND & GN & MF & MP3 & LP & DS & QT & AS & EA & SS & FS & RA & PU & PD \\
        \midrule
        WavMark                      & \textbf{32}    & 4.31            & 39.99           & 8.58   & 0.19   & 1.17   & 3.24   & 0.69   & 1.07   & 1.92   & 1.96   & 0.32   & 0.21   & 2.70   & 3.69   & 5.56   & 49.46   & 47.94 \\
        AudioSeal                           & 16             & 4.28            & 27.53           & 18.66  & \textbf{0.00}     & \textbf{0.00}    & \textbf{0.00}    & \textbf{0.00}    & \textbf{0.00}    & \textbf{0.00}    & \textbf{0.00}    & \textbf{0.00}    & 0.56   & 52.12  & 59.50  & 47.09  & 50.69   & 51.25 \\
        WavMark-16bps                       & 16             & 4.29            & 41.56           & 2.43   & 0.54   & 1.14   & 2.52   & 0.84   & 0.78   & 1.26   & 1.32   & 0.54   & 0.60   & 3.25   & 3.06   & 9.56   & 5.23   & 6.49 \\
        Teacher model                     & 16             & \textbf{4.33}   & \textbf{45.75}  & 0.56   & \textbf{0.00}   & 0.38   & 0.46   & 0.13   & 0.07   & 0.85   & 0.13   & \textbf{0.00}   & \textbf{0.00}   & 1.11   & 1.20   & \textbf{1.45}   & 1.04   & 0.98 \\
        Teacher-woMSSL                   & 16             & 4.29            & 40.34           & 1.23   & \textbf{0.00}   & 0.76   & 2.16   & 0.18   & 0.16   & 1.24   & 0.24   & 0.00   & 0.12   & 2.52   & 2.04   & 6.79   & 2.52   & 2.76 \\
        Teacher-woCMFM                & 16             & 4.30            & 42.25           & 1.77   & 0.60   & 0.84   & 1.68   & 0.84   & 0.78   & 0.96   & 0.90   & 0.54   & 1.02   & 1.86   & 1.86   & 7.03   & 3.43   & 4.45 \\
        
        DKD Student                         & 16             & 4.08            & 44.29           & 1.94   & 0.00   & 2.41   & 1.50   & 0.91   & 2.93   & 6.38   & 0.39   & 0.07   & 0.20   & 3.13   & 3.19   & 2.93   & 4.10   & 2.54 \\
        PKDMark                    & 16             & 4.30            & 45.54           & \textbf{0.51}   & \textbf{0.00}   & 0.39   & 0.58   & 0.19   & 0.32   & 1.23   & \textbf{0.00}   & \textbf{0.00}   & \textbf{0.00}   & \textbf{0.52}   & \textbf{0.85}   & 1.58   & \textbf{0.65}   & \textbf{0.84} \\
        \bottomrule
    \end{tabular}%
    }
\end{table*}
\subsubsection{Joint optimization strategy}
The student model adopts the same attack layer, discriminator, message feature mapping, and adversarial loss as the teacher model. 
The main points of the joint optimization strategy are as follows:
\begin{itemize}[left=0pt, itemindent=0pt, labelsep=0.5em, listparindent=0pt, align=left, wide=0pt, nosep]

    \item \textbf{Perceptual Loss:}
    Different from the first stage, the student perceptual loss is computed as the MSE between the watermark speech of student model $x_w^S(t)$ and the watermark speech of teacher model $x_w^T(t)$. This alignment ensures that the student model’s watermark embedding closely follows the teacher model’s representation in the time-frequency domain, enhancing perceptual similarity and embedding consistency.
\begin{equation}
\mathcal{L}_{\text{per}}^S = \sum_{i \in e} \alpha_i \|STFT_i(x_w^S(t)) - STFT_i(x_w^T(t))\|_2
\end{equation}

    \item \textbf{Decoding Loss:} 
    During joint optimization, the teacher decoder is utilized to decode the messages from the combined output $x_{\text{com}}(t)$ and compute the decoding loss. The MSE loss of encoder and decoder message feature map in first stage is not included during the distillation process.
    \begin{equation}
\mathcal{L}_{\text{dec}}^S = \| m^S - \hat{m}^S \|_2
\end{equation}
    \item \textbf{Adversarial Loss:}
The discriminator of the first stage is utilized in the second stage to provide the adversarial loss of student encoder $\mathcal{L}_{adv}^S$. 
    \begin{equation}
    \mathcal{L}_{adv}^S = \log(1 - d(x_w^S(t)))
    \end{equation}
    \item \textbf{Optimization Strategy:} 
    As in the first stage, the total loss $\mathcal{L_\text{total}^S}$ is computed as a weighted sum of the above three terms. 
    \begin{equation}
    \mathcal{L_\text{total}^S} = \lambda_{\text{per}} \mathcal{L}_{\text{per}}^S + \lambda_{\text{dec}}\mathcal{L}_{\text{dec}}^S + \lambda_{\text{adv}} \mathcal{L}_{\text{adv}}^S \text{,}
\end{equation}
    To maintain training stability, the teacher encoder and decoder’s parameters are initially kept fixed, with updates applied only to the student encoder. Once the student model gradually aligns with the teacher's performance as $\lambda(n)$ reaches 1, a final global fine-tuning step is performed. This approach effectively leverages knowledge from both the teacher encoder and decoder, allowing the student model to converge efficiently while preserving speech quality.
\end{itemize}

\section{Experiments}
\subsection{Experimental setup}
\begin{itemize}[left=0pt, itemindent=0pt, labelsep=0.5em, listparindent=0pt, align=left, wide=0pt, nosep]
\item \textbf{Datasets}.
The experiments are conducted using a composite speech dataset totaling 1,000 hours. It comprises 400 hours from CommonVoice \cite{ardila2020common}, 300 hours from LibriTTS \cite{zen2019libritts}, 200 hours from ASVspoof 2021 DF \cite{yamagishi2021asvspoof}, and 100 hours of high-quality synthetic speech generated via Microsoft Azure Text-to-Speech. This combination provides diverse, natural, and multilingual speech, while ensuring applicability to both real-world speech and DeepFake detection scenarios. The dataset is split into 900/50/50 hours for training/validation/testing. All speech samples are processed at a 24 kHz sampling rate. The STFT and ISTFT operations are performed with an FFT size of 1,024 samples and a hop length of 600 samples.
\item \textbf{Implementation}.
The PKD framework begins with training a teacher model, followed by iterative PKD training to obtain a compact student model. The embedding capacity is set to 16 BPS, which is sufficient for speech synthesis applications. The teacher model comprises 8 INN blocks, each containing a 5-layer ResNet module with 32 channels per layer, while the student model is significantly smaller, with 2 INN blocks and 16 channels per layer. During PKD training, the mixing factor $\lambda(n)$ is linearly increased from 0.1 to 1 over 40k steps.

\item \textbf{Evaluation Metrics}.
Imperceptibility and robustness are assessed using both objective and subjective metrics. For imperceptibility, signal-to-noise ratio (SNR) quantifies watermarking distortion, while perceptual evaluation of speech quality (PESQ) \cite{rix2001perceptual} measures perceptual quality. Additionally, the subjective comparison mean opinion score (CMOS) captures human perception of quality differences. Robustness is evaluated using bit error rate (BER) to assess the reliability of extracted information and detection F1 score to measure accuracy in distinguishing between original and watermarked speech.
\end{itemize}
\subsection{Imperceptibility}
To evaluate imperceptibility, synthesized speech samples from both the teacher and student models are compared against the original recordings. PESQ scores reported in Table~\ref{tab:comparison} indicate minimal quality differences between the teacher (4.33) and student (4.30) models. Furthermore, a CMOS test yields a score of -0.04 when comparing original recordings with student model samples, suggesting that PKDMark is nearly indistinguishable from the original speech. These results confirm that the distillation process effectively preserves high imperceptibility.

\subsection{Robustness}

The robustness is evaluated under various distortions, as summarized in Table~\ref{tab:comparison}.  The tested distortions include: 1) no distortion (ND); 2) Gaussian noise (GN); 3) median filtering (MF); 4) MP3 compression at 64kbps (MP3); 5) low pass filtering with 4kHz stop band (LP); 6) down sampling to 8kHz (DS);   7) quantization to 9bit (QT); 8) amplitude scaling (AS); 9) echo addition with 100ms delay (EA); 10) slow speed to 0.9x (SS) ; 11) fast speed to 1.1x (FS);   12) reverberation attack with RT60 = 200ms (RA); 13) pitch up shifting by +10\% (PU); 14) pitch down shifting by -10\% (PD). Distortions 1) to 9) are basic distortions, while 10) to 14) are advanced attacks, which pose challenges even for some deep learning-based watermark methods\cite{wu2025comparative} \cite{ozer2025comprehensive}. Two open-source models are selected for comparison: WavMark \cite{chen2023wavmark} and AudioSeal \cite{roman2024proactive}. To make a fair comparison, WavMark-16bps is trained at a 16bps bit rate, with the remaining settings kept the same as the original WavMark.

The results presented in Table~\ref{tab:comparison} demonstrate that PKDMark achieves an average BER of 0.51\% across all various distortions, which is comparable to the teacher model's average BER of 0.56\%. Furthermore, PKDMark exhibits superior robustness compared to WavMark and WavMark-16bps across all test distortions, especially under advanced attacks. In particular, PKDMark consistently exceeds WavMark (8. 58\%) and WavMark-16bps (2. 43\%) in all types of distortions.

Compared to AudioSeal, PKDMark achieves significantly better robustness under advanced distortions. While AudioSeal obtains perfect BERs (0.00\%) in several basic distortions, its performance degrades drastically under more complex attacks such as time stretching and pitch shifting, resulting in a high average BER of 18.66\%. In contrast, PKDMark maintains consistently low BERs in most distortion scenarios, highlighting its strong generalization capability under challenging conditions.

To further evaluate watermark reliability, watermark speech detection is performed on a test dataset of 1000 TTS samples, each lasting between 1 and 2 seconds and embedded with an 8-bit synchronization code from the 16-bit capacity. The same distortions in Table~\ref{tab:comparison} are applied. A simple detection strategy is used: if at least 7 out of 8 bits match, the sample is classified as watermarked speech; otherwise, it is classified as original speech. The average detection F1 score reaches 99.6\% across test distortions. 
These results confirm the robustness of our approach in preserving critical watermark message and ensuring reliable detection even under advanced manipulations.

\subsection{Efficiency}
Table~\ref{tab:efficiency} provides a summary of the efficiency gains achieved by PKDMark in terms of computational cost and real-time performance. The computational cost is measured in floating point operations per second (FLOPS). The teacher model requires 36.0 GFLOPS, whereas PKDMark reduces this to 2.3 GFLOPS, resulting in a 93.6\% reduction. Real-time performance is evaluated across both GPU and CPU configurations. All evaluations are conducted using sequential single-threaded processing, without leveraging parallel or multi-threaded inference. On an NVIDIA T4 GPU, PKDMark achieves a real-time factor (RTF) of 0.006, representing an 84.7\% improvement over the teacher model's RTF of 0.039. On an Intel i9-13900 CPU, PKDMark achieves an RTF of 0.020, marking an 87.1\% improvement compared to the teacher model's 0.155. These results highlight the efficiency and practical deployment potential of PKDMark for real-time speech synthesis across diverse hardware environments.
\subsection{Ablation study}
The effectiveness of the proposed components is assessed through a series of ablation experiments, with results summarized in Table 1. 
For the teacher model, Teacher-woCMFM removes the proposed complex message feature map from the teacher model, resulting in an increased BER of 1.77\% and reduced SNR of 42.25 dB. Teacher-woMSSL, which removes the multi-scale STFT loss, yields an increased BER of 1.23\% and reduced SNR of 40.34 dB. These findings demonstrate that both the proposed complex feature map and multi-objective joint optimization are essential to improve the robustness and quality of the watermark.

For the distillation strategy, PKD Student (PKDMark) and DKD Student are trained under identical configurations, differing only in the distillation method. As shown in Table 1, PKD Student achieves superior performance in PESQ (4.30 vs. 4.08) and BER (0.51\% vs. 1.94\%). These results suggest that the PKD strategy enables more effective knowledge transfer, leading to a better balance between perceptual quality and watermark robustness.
\begin{table}[t]
    \centering
    \caption{Computational cost and real-time performance of PKDMark.}
    \label{tab:efficiency}
    \begin{tabular}{lcc*{2}{c}}
        \toprule
        Model & GFLOPS & \multicolumn{2}{c}{RTF ($\downarrow$)} \\
              &       & T4 (GPU) & i9-13900 (CPU) \\
        \midrule
        Teacher Model & 36.0 & 0.039 & 0.155 \\
        PKDMark       & 2.3  & 0.006 & 0.020 \\
        \midrule
        Reduction (\%) & 93.6\% & 84.7\% & 87.1\% \\
        \bottomrule
    \end{tabular}
\end{table}
\section{Conclusions}
In this paper, we propose PKDMark, an efficient speech watermarking method for real-time speech synthesis via progressive knowledge distillation (PKD). Our approach enhances an invertible neural network (INN)-based teacher model to embed watermarks that preserve high fidelity and demonstrate robustness against 14 types of distortions and attacks. By transferring watermarking knowledge from the teacher model to a compact student model, we achieve a significant reduction of 93.6\% in computational cost. Ablation studies further confirm that the proposed enhancements contribute to superior performance relative to the baseline model, with PKD outperforming the DKD approach in both perceptual quality and robustness. These results demonstrate the effectiveness of the proposed PKDMark method, showing that it achieves superior robustness and imperceptibility particularly under advanced attacks compared to the two speech watermarking methods evaluated in this study. Future work will focus on expanding watermark capacity and strengthening robustness to more advanced adversarial conditions, such as neural codec compression and neural denoising attacks.

\bibliographystyle{IEEEtran}
\bibliography{mybib}
\end{document}